\documentclass[]{interact}

\usepackage{epstopdf}
\usepackage[caption=false]{subfig}

\usepackage[numbers,sort&compress]{natbib}
\bibpunct[, ]{[}{]}{,}{n}{,}{,}
\makeatletter
\def\NAT@def@citea{\def\@citea{\NAT@separator}}
\makeatother

\theoremstyle{plain}

\theoremstyle{definition}

\theoremstyle{remark}

\begin{document}

\articletype{ARTICLE}

\title{Three-wall piston-cylinder type pressure cell for muon-spin rotation/relaxation experiments.}

\author{
\name{Rustem Khasanov\textsuperscript{a}\thanks{CONTACT Rustem Khasanov. Email: rustem.khasanov@psi.ch},  Ross Urquhart\textsuperscript{b}, Matthias Elender\textsuperscript{a}, and Konstantin Kamenev\textsuperscript{b}}
\affil{\textsuperscript{a}Laboratory for Muon Spin Spectroscopy, Paul Scherrer Institut, CH-5232 Villigen PSI, Switzerland\\ \textsuperscript{b}Centre for Science at Extreme Conditions, University of Edinburgh, Edinburgh, United Kingdom}
}

\maketitle

\begin{abstract}
A three-wall piston-cylinder type high pressure cell for muon-spin rotation/relaxation experiments was designed, manufactured, tested and commissioned. The outer cylinder of the cell body is made from MP35N and the middle and the inner cylinders are made from NiCrAl nonmagnetic alloys. The mechanical design and performance of the pressure cell are evaluated and optimised using finite-element analysis. The  outcomes of the experimental testing closely match the modelling results. The high-pressure cell is shown to reach pressures of up to 3.3 GPa at ambient temperature, corresponding to 3.0 GPa at low temperatures, without irreversible damage.
\end{abstract}


\section{Introduction}\label{sec:introduction}

Pressure, together with  magnetic field and temperature, is an important thermodynamic physical parameter. The pressure tuning allows to establish phase diagrams, study transitions between different phases, identify critical points, and, sometimes, create new materials with different physical properties compared to their ambient pressure variants.
One of the most prominent example reached in high-pressure physics so far, is the discovery of a room temperature superconductivity in hydrate systems at ultrahigh pressures approaching $p\sim200$~GPa \cite{Drozdov_Nature_2019, Snider_Nature_2020}.

In general, the ‘working horses’ of high-pressure research experiments become the piston-cylinder type of pressure cells. Such cells, on the one hand, may reach moderately high pressures (up to 4.6~GPa at ambient temperature, Ref.~\cite{Fujiwara_JPCS_2014}).  On the other hand, the relatively high `pressure volume' of the piston-cylinder assembly allows not only to put inside the cell the sample itself, but use the complete measurement setup as is made, {\it e.g.}, for resistivity \cite{Fujiwara_JPCS_2014, Walker_RSI_1999, Almax_PC}, ac-suspeptibility \cite{Kurita_PhysC_2003}, nuclear magnetic resonance \cite{Fujiwara_RSI_2007},  specific heat \cite{Gati_RSI_2019}, tunneling \cite{Galkin_PSS_1969}, ultrasound \cite{Kepa_RSI_2016}, thermal expansion experiments \cite{Fietz_HPR_2000}, {\it etc}. A reasonably large volume sample chamber and relatively small pressure cell dimensions have also allowed to use of these types of cells at large facility muon and neutron instruments \cite{Khasanov_HPR_2016, Shermadini_HPR_2017, Podlesnyak_HPR_2018, Wang_RSI_2011, Sadykov_JNR_2018, Sadykov_JPCS_2017}.

In order to maximize pressure in piston-cylinder cells, the methods of `thick-walled cylinder', ‘wound cylinder’, ‘compound cylinder’ and ‘autofrettage’ are invented \cite{Eremets_book_1996, Klotz_book_2013}.

A thick-walled cylinder subject to internal pressure will experience higher compressive tangential (hoop) stress in the inner surface than the outer surface. This means very thick cylinders can only hold a pressure up to half the material yield stress \cite{Eremets_book_1996, Klotz_book_2013}.

\begin{figure}[t]
\centering
\includegraphics[width=0.6\linewidth]{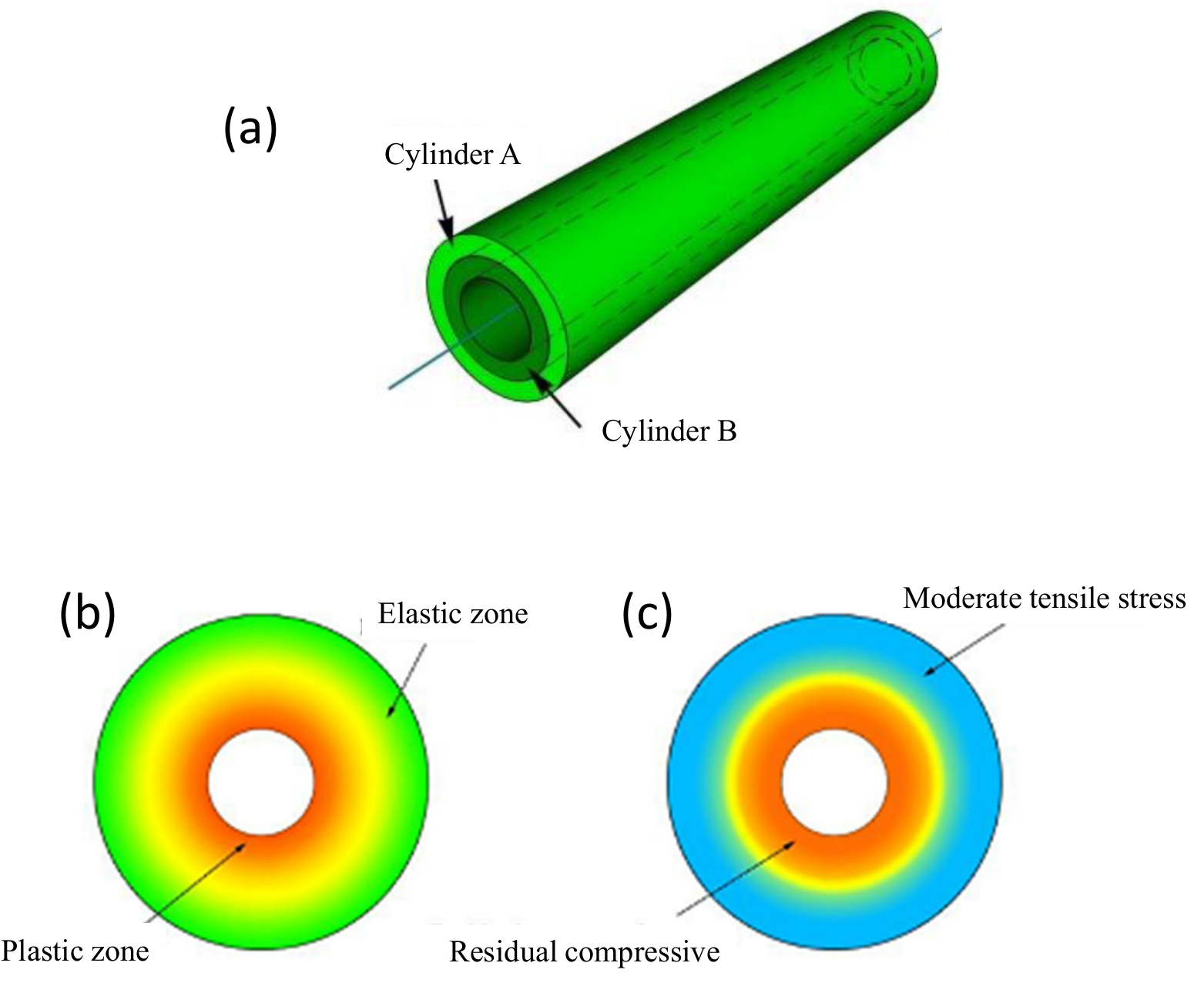}
\caption{(a) Schematic representation of the compound cylinder. The construction is assembled through the shrink- or the press-fit \cite{Khasanov_HPR_2016, Shermadini_HPR_2017, Eremets_book_1996, Klotz_book_2013}. (b) and (c) Schematic representation of the autofrettage process: (b) a large internal pressure produces plastic deformation of the inner pressure cell bore; (c) the residual compressive stresses within the plastic region  left after releasing the applied pressure to zero.  }
\label{fig:compound-cylinder_autofrettage}
\end{figure}

In the wound cylinder technique a strong, highly tensioned belt around a cylinder produces a compressive stress on the core, thus improving the hoop stress distribution across the pressurised cylinder’s wall \cite{Eremets_book_1996, Klotz_book_2013}. Typically, wound cylinders are not used for small piston-cylinder high pressure cells, as cell construction becomes rather complicated.

Compound cylinders are a pre-stressing technique that builds up compressive tangential stress to counteract stress generated through internal pressurisation \cite{Eremets_book_1996, Klotz_book_2013, Abdelsalam_JPVT_2013}. Typically, a two-cylinder construction is used with one cylinder inserted into another with an interference fit, see Fig.~\ref{fig:compound-cylinder_autofrettage}~(a). The inner cylinder experiences compressive tangential stress which counteracts the hoop stress generated through cylinder pressurisation. This decreases the bore hoop stress, allowing the cylinder to withstand higher pressures (see Refs.~\cite{Eremets_book_1996, Klotz_book_2013} and Sec.~\ref{sec:maximum_pressure_in_single-double-triple}). Compound cylinders can be assembled through shrink fit or press fit.

Autofrettage is a cylinder pre-stressing technique that applies controlled internal pressure large enough to produce plastic deformation, Fig.~\ref{fig:compound-cylinder_autofrettage}~(b). Following pressure release, elastic tensile stresses are retained in the outer layers and residual compressive stresses within the plastic region, Fig.~\ref{fig:compound-cylinder_autofrettage}~(c). These compressive stresses counteract hoop stress developed through pressurisation, increasing the allowable cell working pressure \cite{Eremets_book_1996, Klotz_book_2013, Partovi_MasterThesis_2012}.

In this work we have combined the two last techniques, namely the `compound cylinder' and the `autofrettage', in order to produce three-wall pressure cells, {\it i.e.} cells consisting of the one outer and the two inner cylinders.
The main reason for considering the three-wall design, in comparison with the typically used single- or double-wall geometry, is that the maximum reachable pressure in the three-wall assembly is expected to be $\simeq 20$\% higher compared to the double-wall cells and almost double compared to the single-wall cells (see Sec.~\ref{sec:maximum_pressure_in_single-double-triple}). An additional use of the `autofrettage' was expected to strengthen the inner channel of the pressure cell.

Two three-wall pressure cells were built and tested. In both cells, the inner and the two outer cylinders we made out of MP35N and NiCrAl nonmagnetic alloys, respectively.  The pressure in the first cell was increased to the maximum value within the single loading cycle. The failure of the cell, occurred at a maximum pressure $p_{\rm max}\simeq 3.8$~GPa, and was accompanied with the increased plastic deformation of the most inner cylinder. The second cell has obeyed three frettage cycles, with the maximum pressure reached $p_{\rm max}\simeq 3.6$~GPa. This cell was able to maintain pressure $\simeq 3.3$~GPa at a room temperature and $\simeq3.0$~GPa at liquid He temperature.
The mechanical design and performance of the three-wall pressure cells were further evaluated using finite-element analysis (FEA).
In the two identical three-walled piston-cylinder high pressure cell designs, the elastic and plastic deformations were modelled using FEA to find the cell expansion during pressurisation and the resulting plastic deformation at the bore. The simulation and experimental results were very similar. The equivalent stress in each cell at its yield point was also simulated and found to correlate well with the experimental data.

The paper is organized as follows: Section~\ref{sec:maximum_pressure_in_single-double-triple} compares maximum pressures reachable in the single-, double-, and three-wall compound cylinder geometry. Section~\ref{sec:construction_and_FEA_details} describes the  three-wall pressure cell construction, tests of the thread assembly, and details of the finite element analysis procedure. The autofrettage process and the friction inside the cell are discussed in Sec.~\ref{sec:tests_of_the_cell}. Comparison of FEA model simulations and experimental results are made in Sec.~\ref{sec:FEA}. Section~\ref{sec:scientific-example} describes the effect of pressure on the superconducting transition temperature and the thermodynamic critical field of elemental Indium. Conclusions follow in Section~\ref{sec:conclusions}.

\section{Motivation: Maximum pressures for the single-, double-, and three-wall compound cylinders} \label{sec:maximum_pressure_in_single-double-triple}

\begin{figure}[t]
\centering
\includegraphics[width=0.8\linewidth]{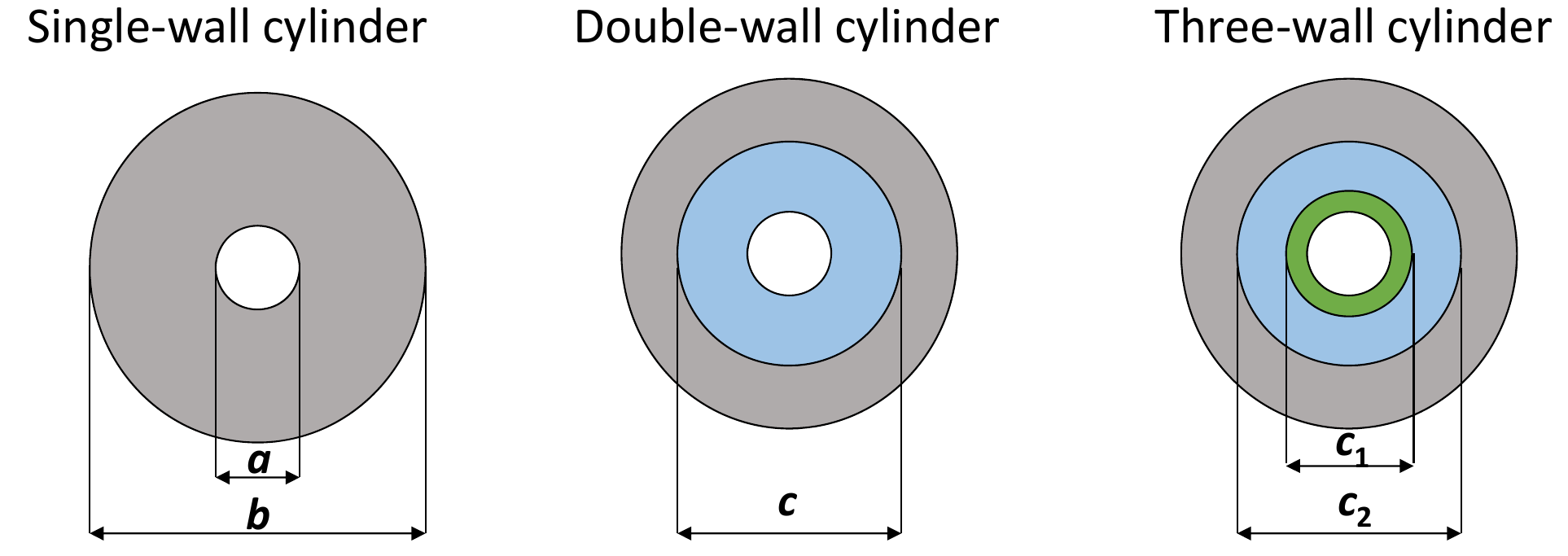}
\caption{The radial cross-section of the single-wall, the double-wall, and the three-wall compound cylinders. $a$ and $b$ are the inner and the outer diameters of the cylinder assembly. $c$, $c_1$, and $c_2$ are the diameters of the inner cylinders.}
\label{fig:single-double-tripple}
\end{figure}

Based on Lam\'{e} equations (see {\it e.g.} Refs.~\cite{Eremets_book_1996, Klotz_book_2013, Shermadini_PhD-Thesis_2014} and references therein) one can easily show that the maximum pressures  for the single-wall ($s$), the double-wall ($d$), and the three-wall ($t$) infinite compound cylinders in the elastic limit are:
\begin{equation}
p_{\rm max}^s=\sigma_{\rm Y}\left[ \frac{1}{2}-\frac{a^2}{2b^2} \right],
\label{eq:p_max-single}
\end{equation}
\begin{equation}
p_{\rm max}^d=\sigma_{\rm Y}\left[ 1-\frac{a^2}{2c^2}-\frac{c^2}{2b^2}\right],
\label{eq:p_max-double}
\end{equation}
and
\begin{equation}
p_{\rm max}^t=\sigma_{\rm Y}\left[ \frac{3}{2}-\frac{a^2}{2c_1^{\;2}}-\frac{c_1^{\;2}}{2c_2^{\;2}}- \frac{c_2^{\;2}}{2b^2}\right].
\label{eq:p_max-triple}
\end{equation}
Here $\sigma_{\rm Y}$ is the yield strength, $a$ and $b$ are the inner and the outer diameters of the cylinder assembly, $c$ is the diameter of the inner cylinder in the double-wall geometry, and  $c_1$ and $c_2$ are the diameters of the first and the second inner cylinders in the three-wall construction (see Fig.~\ref{fig:single-double-tripple}). The optimum values of $c$, $c_1$ and $c_2$, as obtained by minimizing Eqs.~\ref{eq:p_max-double} and \ref{eq:p_max-triple}, are:
\begin{equation}
  c=\sqrt{a b}, \ \ c_1=\sqrt[3]{a^2 b}, \ {\rm and} \ c_2=\sqrt[3]{a b^2}.
\label{eq:c_c1_c2}
\end{equation}
Note that Eqs.~\ref{eq:p_max-single}, \ref{eq:p_max-double}, \ref{eq:p_max-triple} and \ref{eq:c_c1_c2} assume that all cylinders are made of materials with similar yield strengths: $\sigma_{\rm Y,1}=\sigma_{\rm Y,2}=\sigma_{\rm Y,3}=\sigma_{\rm Y}$.

For the pressure cell with $a=6$~mm and $b=24$~mm, which are typical dimensions for the cells used in muon-spin rotation/relaxation experiments \cite{Khasanov_HPR_2016, Shermadini_HPR_2017}, one gets from Eqs.~\ref{eq:p_max-single}, \ref{eq:p_max-double}, \ref{eq:p_max-triple}, and \ref{eq:c_c1_c2} the following ratios for the maximum pressures:
\begin{equation}
p_{\rm max}^s \; \div p_{\rm max}^d \; \div p_{\rm max}^t \simeq 1\; \div 1.6\; \div 1.92.
\label{eq:max-pressure_ratios}
\end{equation}
This suggests that the three-wall assembly allows to almost double the maximum pressure compared to the single-wall design and reaches $\simeq20$\% higher values in comparison to the double-wall geometry.

\section{The construction of the pressure cell and details of the finite element analysis procedure}\label{sec:construction_and_FEA_details}

\subsection{Three-wall pressure cell design and assembly details}

\begin{figure}[t]
\centering
\includegraphics[width=0.7\linewidth]{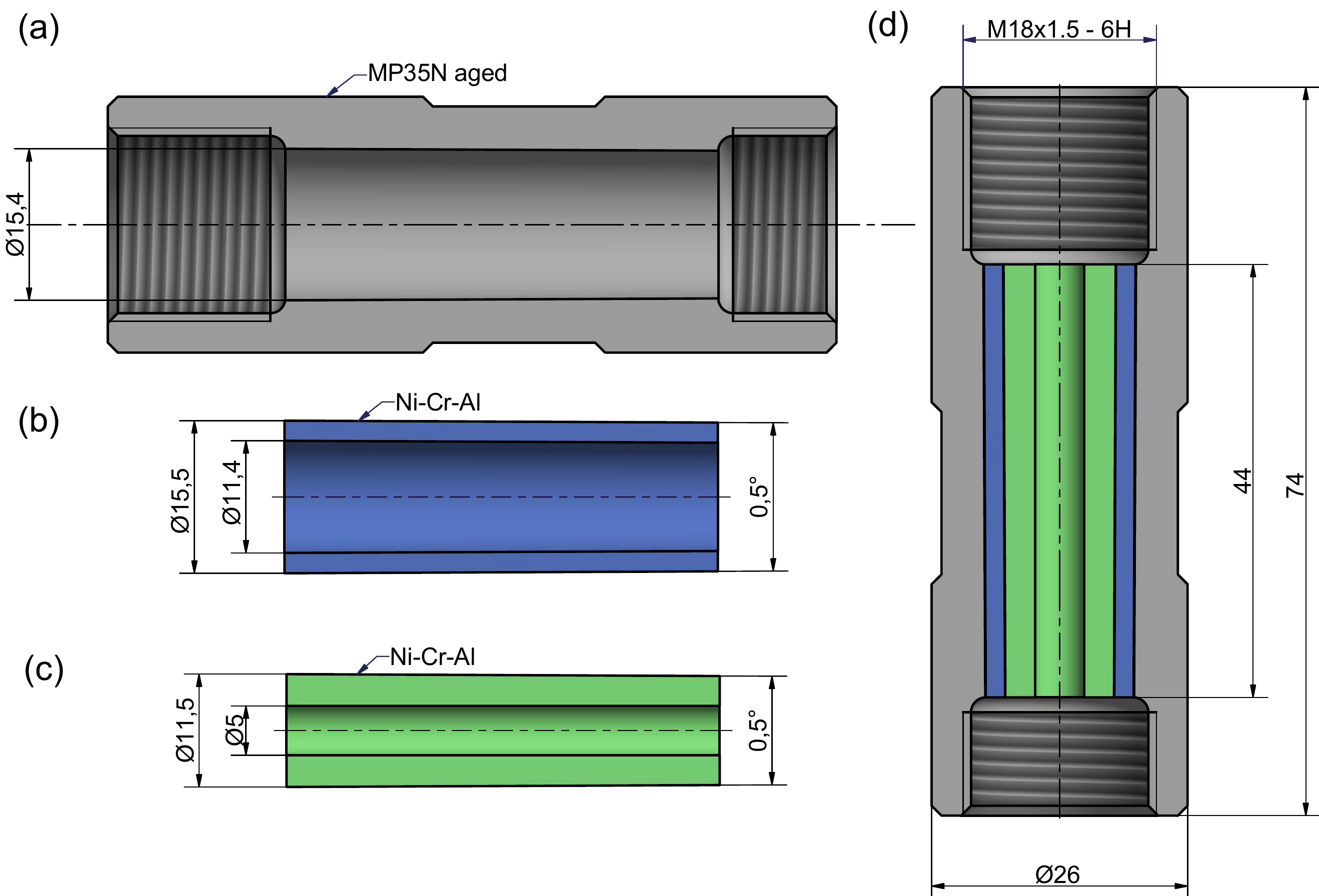}
\caption{Technical drawing of the three-wall pressure cell consisting of: (a) The outer cylinder made out of MP35N alloy; (b) intermerdiate cylinder made of NiCrAl alloy; and (c) inner cylinder made of NiCrAl alloy. (d) Three cylinders assembled together via the press fit. }
\label{fig:three-wall-cell}
\end{figure}

The schematic view of a three-wall piston-cylinder pressure cell is presented in Figs.~\ref{fig:three-wall-cell} and \ref{fig:three-wall-cell_bolts}. Two pressure cells were manufactured and tested from this design. Cell \#1 underwent one pressure cycle until failure. Cell \#2 underwent autofrettage to build residual stress within the cell wall, before being used for an experiment. The inner and middle cylinders were made from NiCrAl alloy with a yield strength of around $\sigma_{\rm Y}^{\rm NiCrAl}\simeq 2.0$~GPa and elastic modulus of $E^{\rm NiCrAl}\simeq270$~GPa \cite{RandD_2009}. The outer cylinder was made from age hardened MP35N alloy with a yield strength of $\sigma_{\rm Y}^{\rm MP35N}\simeq 1.98$~GPa and elastic modulus of $E^{\rm MP35N}\simeq228$~GPa \cite{Shermadini_HPR_2017}.

The pressure cell consists of three cylindrical pieces. The outer MP35N cylinder [Fig.~\ref{fig:three-wall-cell}~(a)] has an inner diameter of 15.4~mm and outer diameter of 24~mm (here after, in a case of conically machined  cylinder surfaces, the highest diameter value is given).  The middle NiCrAl cylinder [Fig.~\ref{fig:three-wall-cell}~(b)] has an inner and outer diameter of 11.4~mm and 15.5~mm, respectively. The inner NiCrAl cylinder [Fig.~\ref{fig:three-wall-cell}~(c)] has a bore diameter of 5.0~mm and an outer diameter of 11.5~mm. The bore of the outer cylinder, the inner and the outer surfaces of the middle cylinder and the outer surface of the inner cylinder are machined conically with the angle $0.5^{\rm o}$.
To strengthen parts of the cell staying outside of the muon beam, the top and the bottom diameters of the outer MP35N
cylinder were increased up to 26~mm [see Fig.~\ref{fig:three-wall-cell}~(a)].

In prior of assembly, the machined cylinders were hardened using
temperature annealing procedure: the outer MP35N cylinder -- 4 hours at $590^{\rm o}$C air cooled;  the middle NiCrAl cylinder -- 2 hours at $700^{\rm o}$C air cooled; and the inner NiCrAl cylinder -- 2 hours at $650^{\rm o}$C for  the cell \#1 and 2 hours at $650^{\rm o}{\rm C}+ 2$~hours at $700^{\rm o}$C  for the cell \#2, respectively.
During the assembly process, the three cylindrical pieces were pressed one inside the other by using hydraulic press. The mechanical drawing of the assembled pressure cell body is shown in Fig.~\ref{fig:three-wall-cell}~(d).

\subsection{Holding forces of the threaded parts of the pressure cell}

\begin{figure}[t]
\centering
\includegraphics[width=0.7\linewidth]{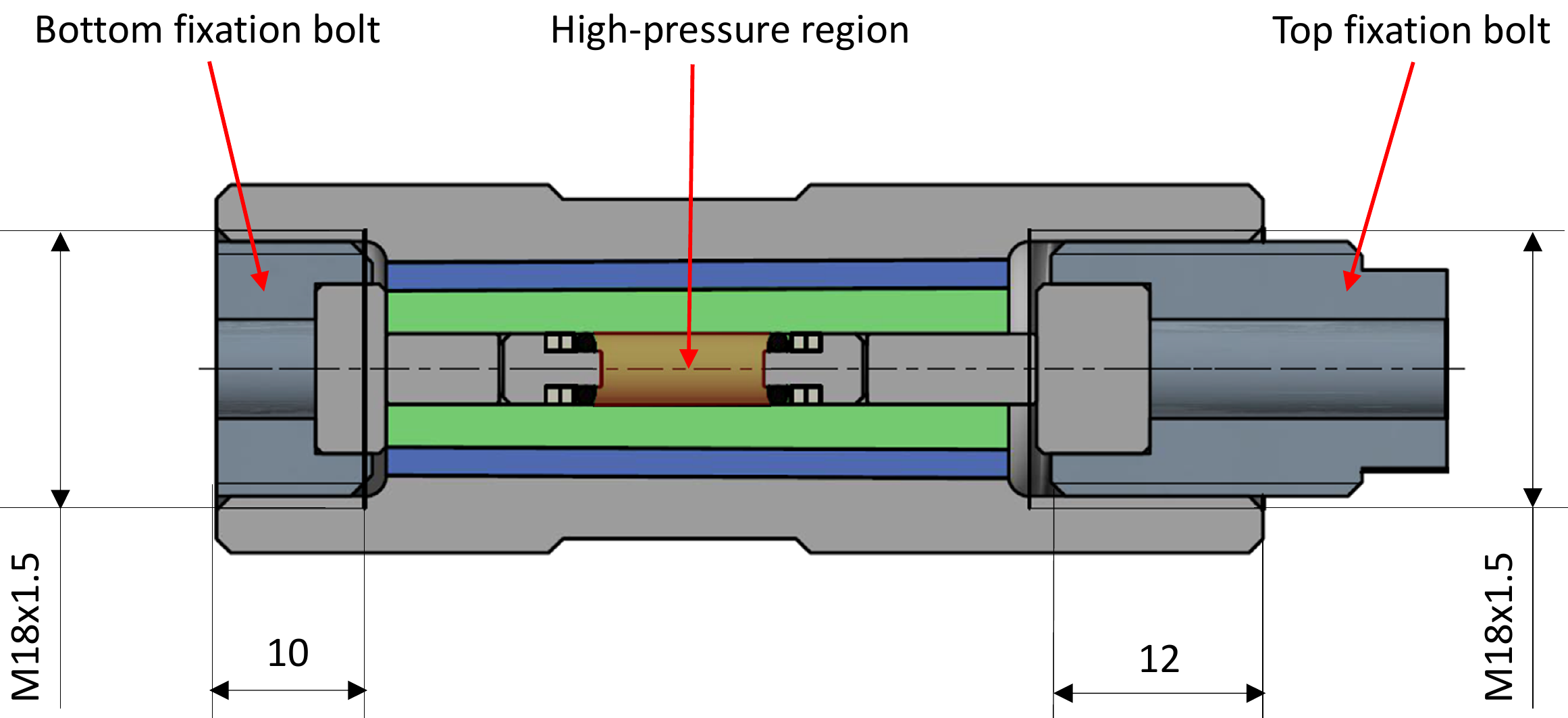}
\caption{The assembled three-wall pressure cell. The top and bottom fixation bolts, used to hold WC pistons at the position, are made out of hardened MP35N alloy. The top and bottom bolts are engaged into  the pressure cell body by $\sim 12$ and $10$~mm, respectively. Note that the position of the 'top bolt` depends on the value of the applied pressure.  }
\label{fig:three-wall-cell_bolts}
\end{figure}

The mechanical drawing of the assembled three-wall pressure cell is shown in Fig.~\ref{fig:three-wall-cell_bolts}.
Additional parts, including the top and the bottom fixation bolts, mushroom seals, pistons, and compressing pads, are the same as described in Refs.~\cite{Khasanov_HPR_2016, Shermadini_HPR_2017} and are not discussed here.
The pressure inside the cell (the orange region between two mushroom seals, Fig.~\ref{fig:three-wall-cell_bolts}) is held  by the pistons, which are supported by the top and bottom fixation bolts. They are made of MP35N alloy, {\it i.e}, on the same material as the outer cell body [see Fig.~\ref{fig:three-wall-cell}~(a)]. The bottom bolt is screwed inside the pressure cell body by $\simeq 10.0$~mm. The position of the top bolt is not fixed and it depends on the loading conditions and the pressure inside the cell. At the highest applied pressure, the top bolt is screwed in by 11 to 14~mm. In Figure~\ref{fig:three-wall-cell_bolts} it enters the cell by 12.0~mm.

Following the assembly drawing presented in Fig.~\ref{fig:three-wall-cell_bolts}, a special attention needs to be paid not only to the central region of the cell where the high-pressure develops, but also to the end parts of the cell where the top and the bottom screws enter the pressure cell body.  In these parts of the cell two possible types of failure may occur: (i) the rupture of the cell body at the threaded part and (ii) the stripping of the body/bolt thread.

The maximum holding force for the case (i) could be estimated as: $F_{\rm max}=S_{\rm body}\;\times\;\sigma_{\rm Y}$. Here $S_{\rm body}=\pi\times(d_{\rm o}^2-d_{\rm i}^2)/4$ is the area of threaded part of the pressure cell body [$d_{\rm o}=26$~mm and $d_{\rm i}=$18~mm are the outer the inner diameters, respectively, Fig.~\ref{fig:three-wall-cell}~(a)]. With $\sigma_{\rm Y}^{\rm MP35N}\simeq 1.98$~GPa \cite{Shermadini_HPR_2017}, one gets $F_{\rm max}\simeq 550$~kN or $\simeq 55$~ton.

As for the second point, the stripping of the bolt/nut thread, the FEA approach is known to result in unreliable outputs. One needs, therefore, to use approaches based on real test experiments. Following Ref.~\cite{Schwarz_2000} the stripping force depends on the material tensile strength, the lengths of the bolt/nut connection, and on the thread diameter. The authors of Refs.~\cite{Schwarz_2000} have proven their calculations by performing mechanical tests on M16 and M20 steel screws, {\it i.e.} on screws with diameters similar to that used in the three-wall construction. By substituting the thread and material parameters for the three-wall cell into the equations derived in Ref.~\cite{Schwarz_2000} one obtaines $\simeq300$ and $\simeq250$~kN stripping of force values for the top and bottom threaded assembly, respectively.
The results of `rupture' and `striping of' calculations suggest, therefore, that the weakest part of the cell stays at the bottom and that the highest force, which threaded parts of the cell could withstand, corresponds to $\simeq250$~kN or $\simeq 25$~ton.

In order to prove these simulations, an additional mechanical test was performed. The 200~kN compressing force was applied to the top fixation bolt screwed by 10~mm inside the pressure cell body. No visible damage was detected on both, the pressure cell body and the fixation bolt, after releasing the force down to zero.

\subsection{Finite element analysis (FEA) model}

The cell was modelled using ANSYS Workbench, Ref.~\cite{ANSYS},  to conduct an FEA study of the cell stress and deformation. The pressure cell area of interest is the cell wall. This can be simply represented as a 2-D axisymmetric model due the symmetrical cell geometry and loading. This allows for a finer mesh size to be used on the model, reducing the computing time to converge upon a solution compared to a 3-D model.

\begin{figure}[t]
\centering
\includegraphics[width=0.5\linewidth]{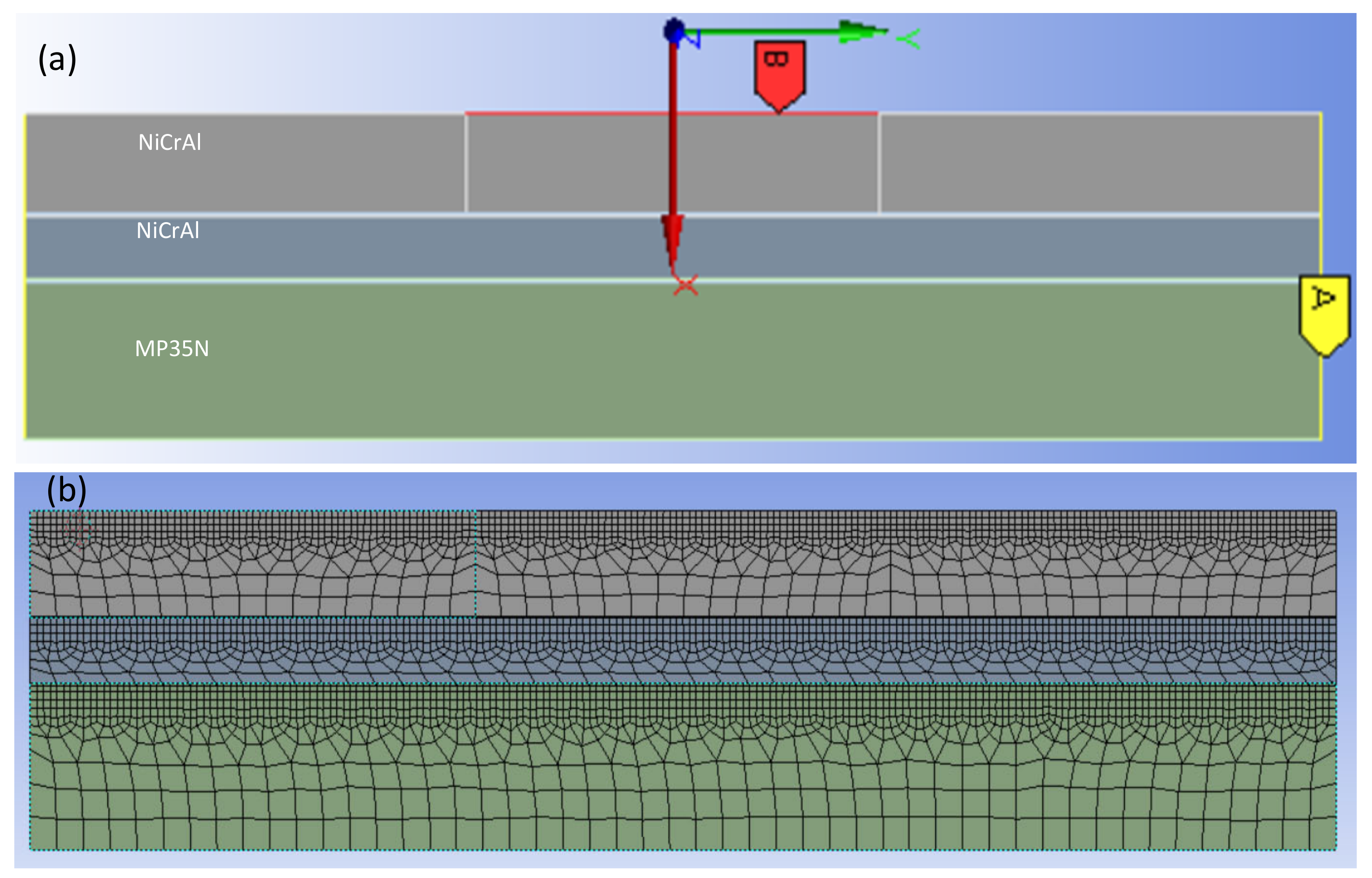}
\caption{(a) FEA model of the pressure cell walls. The boundary condition is the zero displacement in the $y-$direction at each end of 44~mm long cylinder (region A). In the loading condition the pressure is applied to a 14~mm section of the bore (reigon B). (b) The mesh model. In order to improve the accuracy of FEA calculations, a more refined mesh is used in the areas of highest loading.}
\label{fig:FEA-model}
\end{figure}

The model was set up as shown in Fig.~\ref{fig:FEA-model}. A more refined mesh was used in the areas of highest loading, improving result accuracy. To further simplify the model, the threaded section of the pressure cell was not included in the model as it has little effect on the stress and strain in the wall of the pressure cell. The ends of the pressure cell were, therefore, constrained from movement in the $y-$direction, where the fixing bolts would be. The pressure was applied to a 14~mm section of the bore, just under a third of the pressure cell length of 44~mm, for each simulation.

\section{The pressure cell frettage and loading procedure}\label{sec:tests_of_the_cell}

\subsection{The autofrettage procedure}\label{sec:autofretage_procedure}

The autofrettage process is used to increase the maximum reachable pressure value. The physical processes leading to improved mechanical properties of the pressure cell bore are briefly discussed in the introductory part (see Sec.~\ref{sec:introduction}) and presented in Fig.~\ref{fig:compound-cylinder_autofrettage}~(b) and (c). A more detailed description of the autofrettage processes could be found in Refs.~\cite{Shermadini_HPR_2017, Eremets_book_1996, Klotz_book_2013, Abdelsalam_JPVT_2013}.

Mechanically, the autofrettage procedure consists of two steps. At step one, the pressure inside the cell increases until the plastic deformations become detectable. At step two, the applied pressure is released to a certain low value in order to reduce stresses accumulated inside the cell. The steps 'one' and 'two' are then repeated several times.
Due to mechanical hardening of the pressure cell channel, each subsequent pressure application cycle increases the plastic deformation onset point.
By finishing the autofrettage procedure, the pressure as high as the pressure corresponding to the appearance of plastic deformations in the last autofrettage cycle might be applied \cite{Eremets_book_1996, Klotz_book_2013}.

\begin{figure}[t]
\centering
\includegraphics[width=0.6\linewidth]{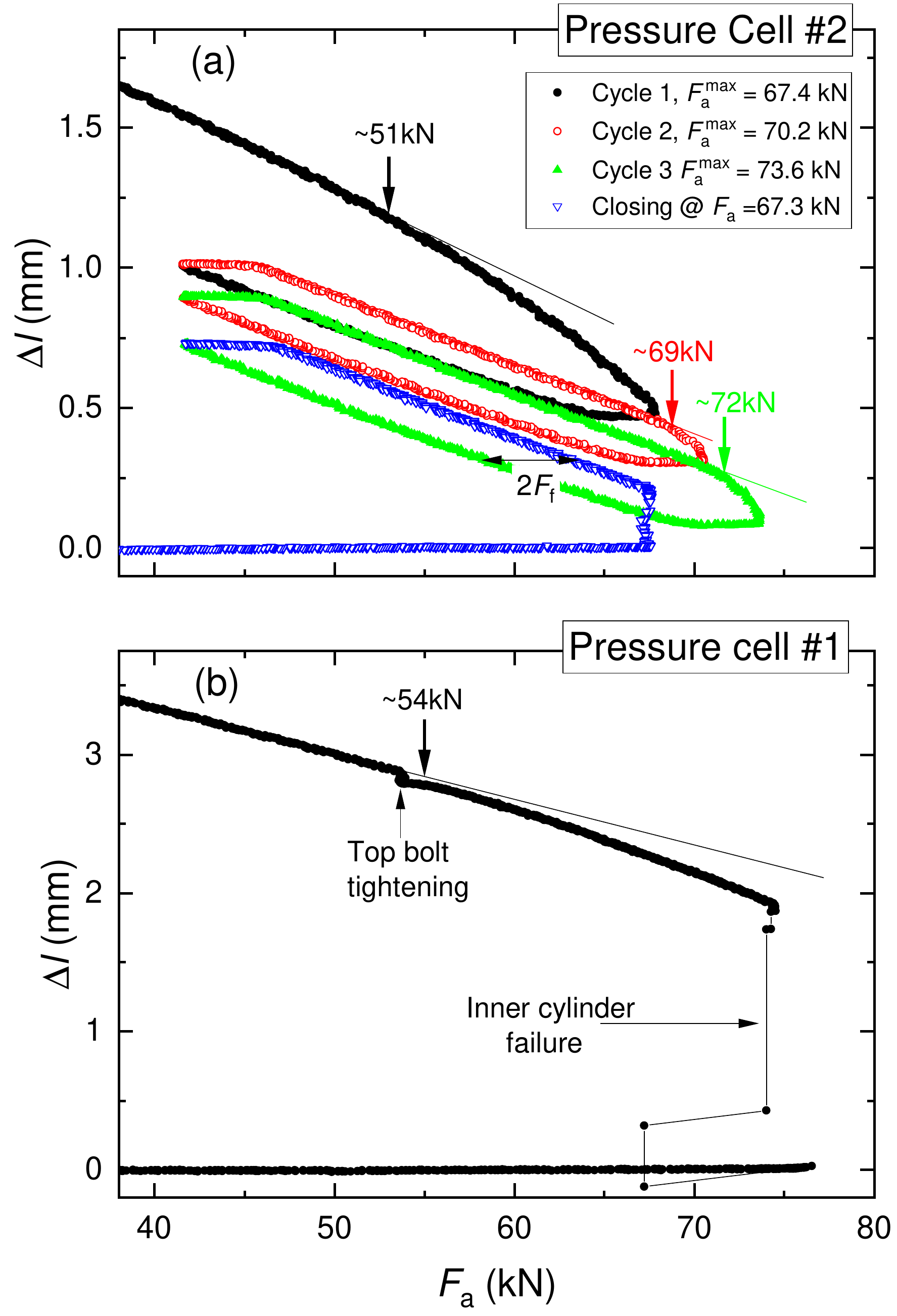}
\caption{(a) The displacement of the piston $\Delta l$ as a function of the applied force $F_{\rm a}$ for the cell \#2. The first three cycles with the maximum applied force $F_{\rm a}^{\rm max} =67.4$~kN (cycle 1, black closed circles), $F_{\rm a}^{\rm max} =70.2$~kN (cycle 2, open red circles) and $F_{\rm a}^{\rm max} =73.6$~kN (cycle 3, closed green up triangles) represent the autofrettage process. Arrows at $\simeq51$, $\simeq 69$ and $\simeq 72$~kN represent the onset of `plastic deformations'. The closing of the cell was performed at $F_{\rm a}^{\rm max} =67.3$~kN (open blue down triangles). By tightening the `top' bolt of the pressure cell (see Fig.~\ref{fig:three-wall-cell}) the piston was moved by an additional $\simeq 0.2$~mm inside the cell. The horizontal double sided arrow represent the width of the hysteresis, which is equal to twice the frictional force $2 F_{\rm f}$. (b) The failure of the cell \#1 which was not obeying the autofrettage. }
\label{fig:autofrettage_cell2}
\end{figure}

\begin{figure}[t]
\centering
\includegraphics[width=0.6\linewidth]{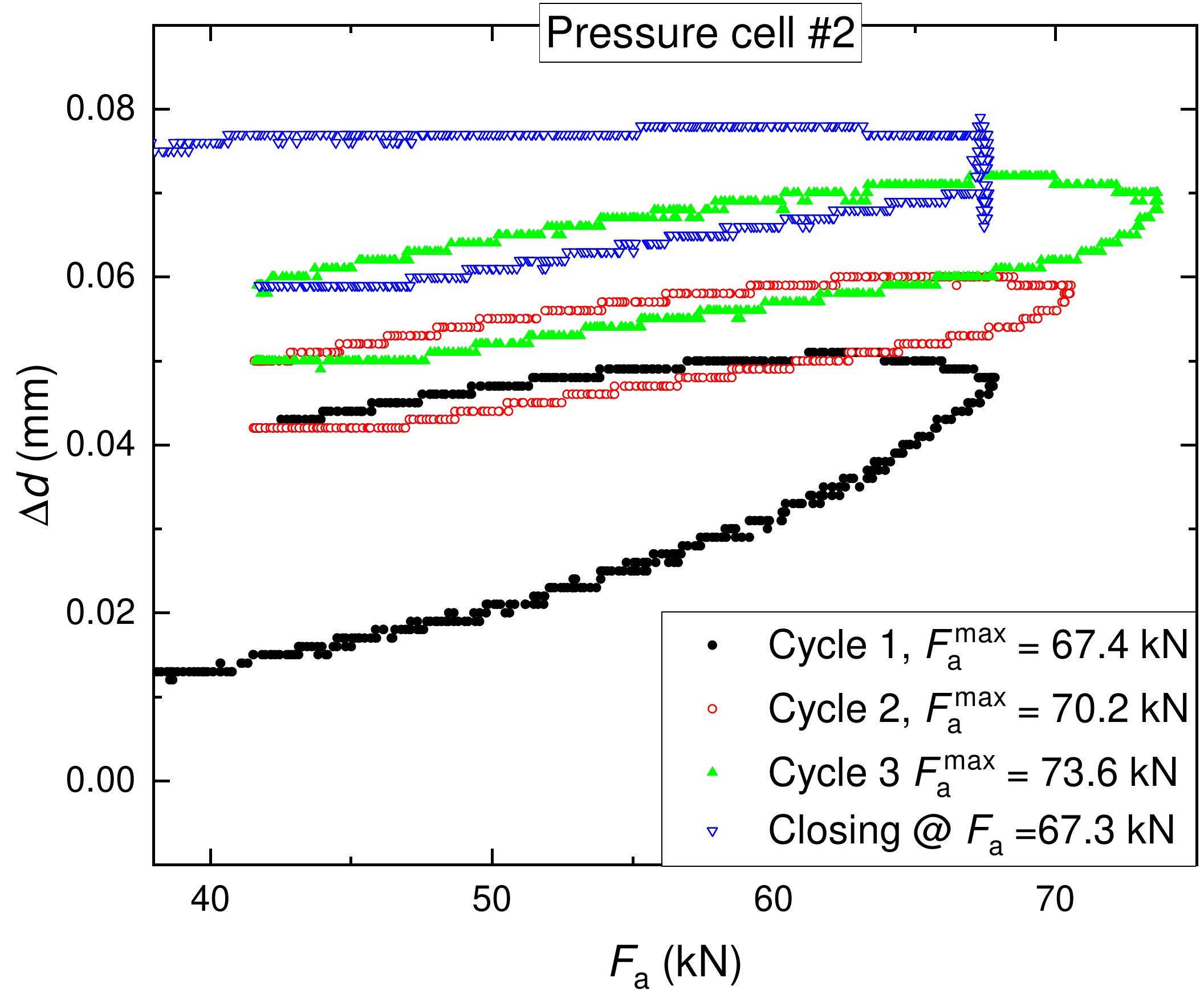}
\caption{The increase of the diameter $\Delta d$ as a function of the applied force $F_{\rm a}$ for the cell \#2. The first three cycles with the maximum applied force $F_{\rm a}^{\rm max} =67.4$~kN (cycle 1, black closed circles), $F_{\rm a}^{\rm max} =70.2$~kN (cycle 2, open red circles) and $F_{\rm a}^{\rm max} =73.6$~kN (cycle 3, closed green up triangles) represent the autofrettage process. Post autofrettage, the cell was closed at $F_{\rm a}^{\rm max} =67.3$~kN (open blue down triangles). }
\label{fig:Radial-expansion_cell2}
\end{figure}

Figure~\ref{fig:autofrettage_cell2}~(a) represents the autofrettage process of the cell \#2. Prior to pressurization, the cell was loaded with the `soft' metal, Indium. The use of the soft matter (like In, Sn, Pb or Teflon) is important, since it allows to ensure that the force, applied via the top piston, is randomly distributed over the pressure cell inner volume. The total height of Indium cylinder inside the cell was $\simeq 15$~mm at ambient pressure.
The pressure cell loading was monitored by using a computer-controlled system allowing to measure in-situ the displacement of the piston ($\Delta l$) and the radial expansion of the pressure cell ($\Delta d$) as a function of the applied force $F_{\rm a}$ (see Figs.~\ref{fig:autofrettage_cell2} and \ref{fig:Radial-expansion_cell2}). The arrows in Fig.~\ref{fig:autofrettage_cell2} indicate the onset of the plastic deformations inside the cell, which are determined via deviations of the loading curves from the linear (elastic) slope.
At each autofrettage cycle the applied force (after reaching the maximum force value $F_{\rm a}^{\rm max} =67.4$~kN in the cycle 1, $70.2$~kN in the cycle 2, and $73.6$~kN in the cycle 3) was released down to $F_{\rm a}\simeq 40.5$~kN. The frettage effect is clearly visible via continuous increase of the onset point of the plastic deformations, which changes from $\simeq 51$~kN in the first cycle to $\simeq 72$~kN in the third cycle [see Fig.~\ref{fig:autofrettage_cell2}~(a)].

The importance of performing the autofrettage procedure, in prior of using the cell in real experiments, is seen from  Fig.~\ref{fig:autofrettage_cell2}~(b), which shows the loading curve of the cell \#1. This cell was pressurised within the single cycle until the failure of the weakest part, the inner NiCrAl cylinder, occurred.

\subsection{Friction inside the pressure cell}

\begin{table}[htb]
\centering
\caption{\label{Table1} Applied force and true cell pressure in cell \#2. The meaning of the parameters is: $F_{\rm a}$ is the applied force, $F_{\rm a}-F_{\rm f}$ is the force without friction, and $p_{\rm t}$ is the true pressure inside the cell.\\}
\begin{tabular}{cccccc}
\hline
\hline
Cycle&$F_{\rm a}$&$F_{\rm a}-F_{\rm f}$&$p_{\rm t}$&\\
           & (kN)       &(kN)          &(Gpa)\\
\hline
1&67.4&64.9&3.31\\
2&70.2&67.7&3.45\\
3&73.6&71.1&3.62\\
Closing the cell&67.3&64.8&3.30\\
\hline
Depressurisation after&40.5&38&1.90\\
each cycle&&&\\
\hline
\hline
\end{tabular}
\end{table}

The force $F_{\rm a}$, applied by means of the external press to the piston,  could be converted into a pressure using:
\begin{equation}
  p=\frac{F_{\rm a}}{S_{\rm bore}}.
\label{eq:pressure}
\end{equation}
Here $S_{\rm bore}$ is the bore area. To ensure the pressure cell does not leak, a good sealing system is needed. The tight seal required for the piston-cylinder type of pressure cells can lead to a large amount of friction due to the piston sliding past the cell wall during pressurisation/depressurisation \cite{Walker_RSI_1999}. This friction is the reason for the hysteresis curves that occur between the loading and unloading autofrettage steps [see Fig.~\ref{fig:autofrettage_cell2}~(a)]. Before the applied force acts to pressurise the cell, frictional forces must be overcome. Therefore, the true pressure in the cell ($p_{\rm t}$) is less than the applied pressure during cell loading ($p_{\rm t}<F_{\rm a}/S_{\rm bore}$) and greater during unloading ($p_{\rm t}>F_{\rm a}/S_{\rm bore}$). Consequently, the width of the hysteresis curve corresponds to twice the frictional force ($2 F_{\rm f}$) of the sealing system. The true pressure inside the cell could be found, therefore,  as:
\begin{equation}
  p_{\rm t}=\frac{F_{\rm a}-F_{\rm f}}{S_{\rm bore}},
\end{equation}
with $F_{\rm f}$ equals half the width of the hysteresis curve [$\simeq 2.5$~kN, see Fig.~\ref{fig:autofrettage_cell2}~(a)]. Table~\ref{Table1} shows the true pressure in cell \#2 for each loading cycle.

\section{Finite element analysis}\label{sec:FEA}

\subsection{FEA verification}

Experimental pressure cell data allows for verification of the ANSYS FEA model through comparison of the simulation and experimental results. Figure~\ref{fig:Radial-expansion_cell2} shows the pressurisation of the cell \#2 in terms of the cell outer radial expansion $\Delta d$. Post autofrettage, the cell was closed at $p_{\rm t}=3.3$~GPa [see also Fig.~\ref{fig:autofrettage_cell2}~(a) and Table ~\ref{Table1}].

It is necessary to model both the plastic and elastic deformation in ANSYS within the pressure cell wall. True stress-strain considers the actual area of the test specimen, providing more accurate material properties during plastic deformation. The engineering stress-strain curves for NiCrAl and MP35N, Refs.~\cite{Shermadini_HPR_2017, RandD_2009}, were therefore converted into true stress-strain using the relations:
$\sigma_{\rm T}=\sigma_{\rm e} (1+\epsilon_{\rm e})$ and $\epsilon_{\rm T}=ln(1+\epsilon_{\rm e})$ \cite{Choudhary_MMT_2013}. Here $\sigma_{\rm e}$/$\sigma_{\rm T}$ and $\epsilon_{\rm e}$/$\epsilon_{\rm T}$ are the engineering/true stress and strain, respectively.  The material bilinear isotropic hardening properties for the ANSYS model were found from the tangent modulus of the plastic deformation region of the true stress-strain curves and were $\simeq22$~GPa and $\simeq40$~GPa for NiCrAl and MP35N, respectively.

\begin{figure}[t]
\centering
\includegraphics[width=0.6\linewidth]{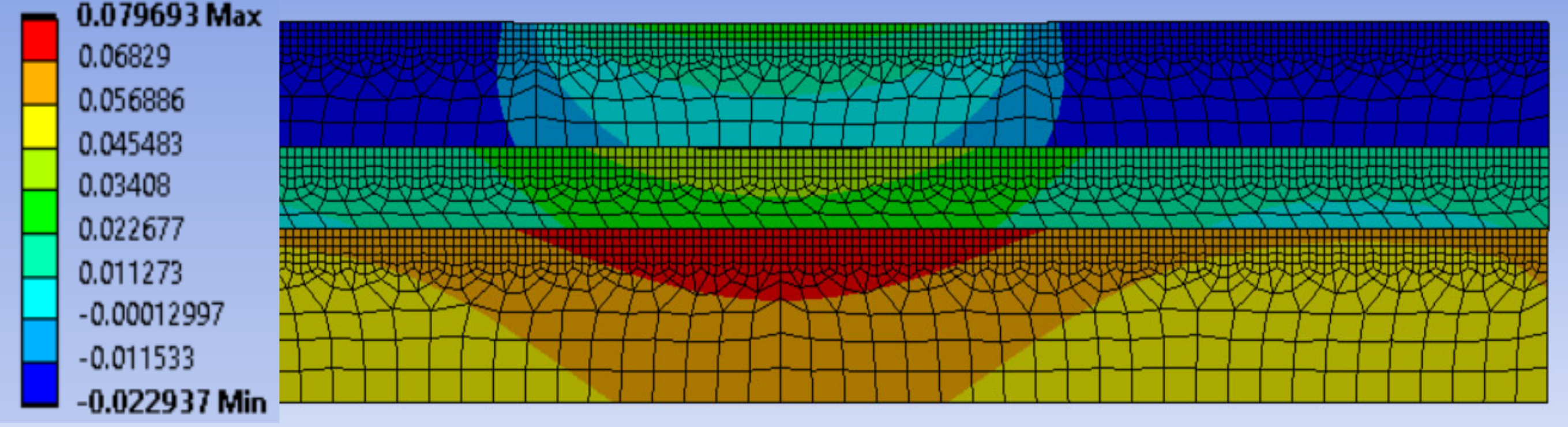}
\caption{FEA results for total deformation in the three-wall pressure cell at pressure $p=3.3$~GPa. }
\label{fig:Radial-expansion_ANSYS}
\end{figure}

From Figure~\ref{fig:Radial-expansion_cell2} the pressure cell radial expansion between zero pressure and cell close at 3.3~GPa is approximately 0.07~mm. The total cell combined elastic and plastic deformation modelled in the ANSYS FEA was calculated to be $\simeq 0.06$~mm at the cell outer radius (see Fig.~\ref{fig:Radial-expansion_ANSYS}). This is $\simeq0.01$~mm less than the experimental value.

The plastic deformation was also calculated at the bore after the autofrettage process was complete and the cell was unloaded. The effect of loading and unloading cycles are not significant on the total bore deformation due to strain hardening when material yield strength is exceeded. Therefore, the analysis is simplified by finding the plastic strain at maximum pressure of 3.62~GPa [see Fig.~\ref{fig:compound-cylinder_autofrettage}~(a), Fig.~\ref{fig:Radial-expansion_cell2}, and Table~\ref{Table1}] which is used to calculate the change in the inner bore diameter.
The simulation results in total diametral deformation of $\simeq0.08$~mm. After autofrettage, the cell \#2 was depressurised, and a micrometre was used to re-measure the bore, which was found to increase by 0.09~mm to 0.1~mm due to plastic deformation. This is between 0.01-0.02~mm more than the FEA simulation results.

It should noted here that the plastic deformation at the bore and the total radial deformation are conservative compared to experimental results. However, the FEA and the experiment for both radial expansion and bore deformation show close correlation, proving the validity of the ANSYS model.

\subsection{Stress analysis}

Cell \#1 did not undergo autofrettage and was pressurised to failure, which occurred in the inner NiCrAl cylinder. Cell \#2 underwent autofrettage and was not pressurised to failure. The onset of plastic deformations for the cell \#1 and \#2 are presented in Fig.~\ref{fig:autofrettage_cell2}. Table~\ref{Table2} shows the true internal pressures generated by the applied force at the cell \#1 fail and in both cells at the onset point of the first obtained plastic deformation. The stress in the cell \#1 and cell \#2 at their yield point were analysed by applying the respective true pressure shown in Table~\ref{Table2} to the ANSYS model. The results presented below are given in the units of the equivalent stress \cite{Klotz_book_2013, Eremets_book_1996, Shermadini_PhD-Thesis_2014}:
\begin{equation}
\sigma_{\rm eq}=\sigma_{\rm t}-\sigma_{\rm r},
 \label{eq:equivalent_stress}
\end{equation}
with $\sigma_r$ and $\sigma_t$ being the radial and the tangential stress, respectively.

\begin{table}[htb]
\centering
\caption{\label{Table2} Applied force and corresponding true pressure for the key areas for analysis in each cell.\\ }
\begin{tabular}{cccccc}
\hline
\hline
Cell&$F_{\rm a}$&$F_{\rm a}-F_{\rm f}$&$p_{\rm t}$&\\
           & (kN)       &(kN)          &(Gpa)\\
\hline
\#1: Fail&77&74.5&3.79\\
\#1: Yield&54&51.5&2.62\\
\#2: First Yield&51&48.5&2.47\\
\hline
\hline
\end{tabular}
\end{table}

\begin{figure}[t]
\centering
\includegraphics[width=0.7\linewidth]{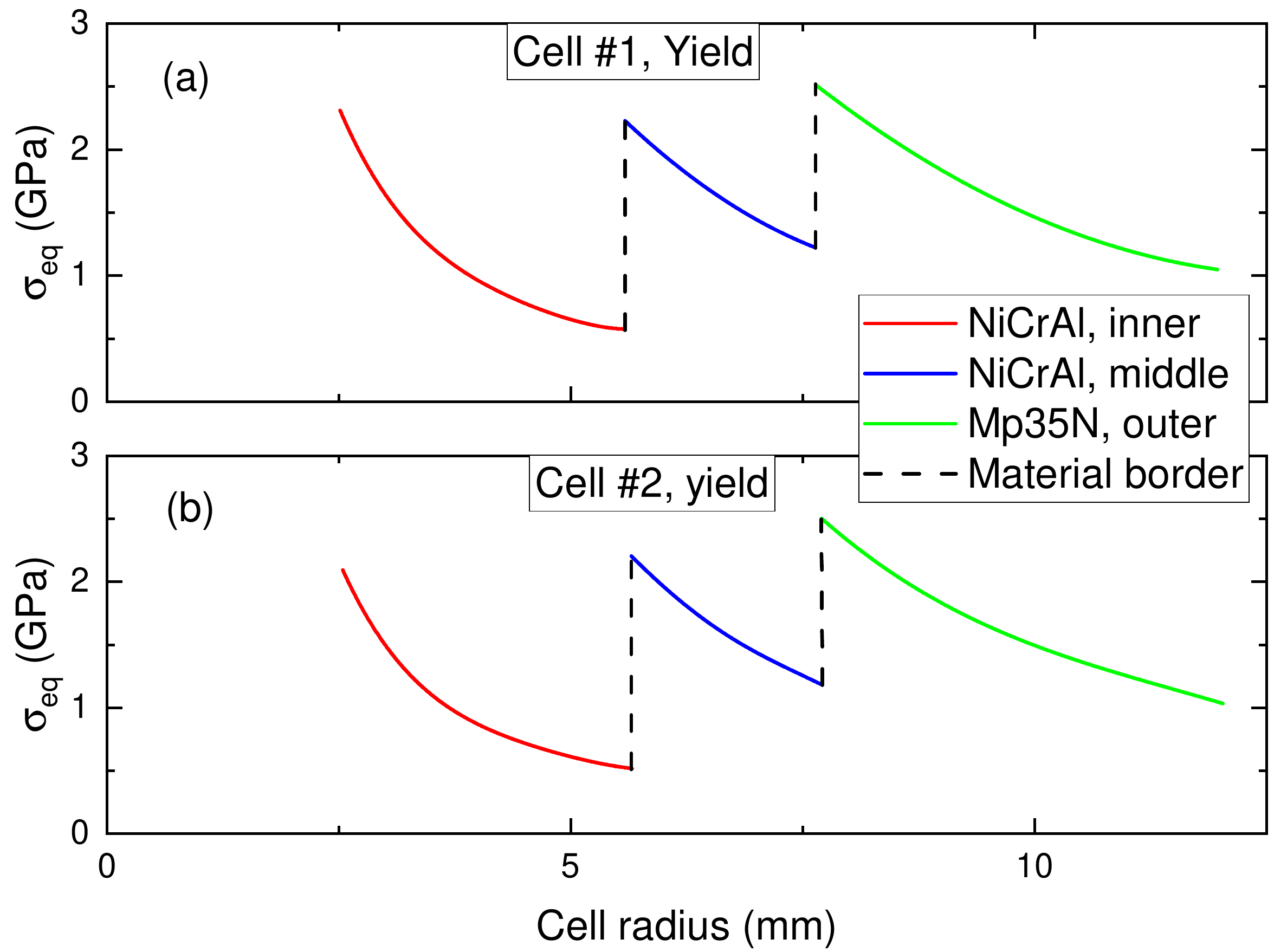}
\caption{(a) Equivalent stress distribution in the cell \#1 at $p=2.62$~GPa. The maximum equivalent stress in the outer cylinder is around 2.5~GPa, which is $\sim0.5$~GPa over the MP35N yield strength ($\sigma_{\rm Y}^{\rm MP35N}\simeq 1.98$~GPa, Ref.~\cite{Shermadini_HPR_2017}). The middle and the inner cylinders have a maximum stress $\simeq 0.1$ and $\simeq 0.4$ above the NiCrAl yield strength ($\sigma_{\rm Y}^{\rm NiCrAl}\simeq 2.0$~GPa, Ref.~\cite{RandD_2009}), respectively. (b) The same is in panel (a) but for the cell \#2 at $p=2.47$~GPa. The maximum equivalent stresses are $\simeq 2.5$, 2.2, and 2.1~GPa for the outer, middle, and inner cylinders, respectively.}
\label{fig:Equalent-Stresses_cell_12}
\end{figure}

From Table~\ref{Table2} the yield pressure of the cell \#1 was 0.15~GPa higher than for the cell \#2. The FEA model was modelled as fully elastic to allow for a clearer representation of equivalent stress data, which would not be influenced by variations in material yield strength. The equivalent stress
results from the FEA for the cell \#1 and  \#2 are presented in Fig.~\ref{fig:Equalent-Stresses_cell_12}.

Following Figure~\ref{fig:Equalent-Stresses_cell_12}, the maximum equivalent stress
is approximately 2.2~GPa in the middle NiCrAl cylinder and 2.5~GPa in the outer MP35N cylinder for both cells: \#1 at $p=2.62$~GPa and \#2 at $p=2.47$~GPa. The middle cylinder has a maximum stress slightly above NiCrAl yield strength ($\sigma_{\rm Y}^{\rm NiCrAl}\simeq 2.0$~GPa, Ref.~\cite{RandD_2009}). The maximum stress in the outer cylinder is around 0.5~GPa over the MP35N yield strength ($\sigma_{\rm Y}^{\rm MP35N}\simeq 1.98$~GPa, Ref.~\cite{Shermadini_HPR_2017}), falling below yield strength in less than 1~mm into its radius.

The main difference occurs for the inner NiCrAl cylinders, where the maximum equivalent stress  is $\simeq2.4$~GPa in the cell \#1, which is $\simeq0.3$~GPa larger than in the cell \#2 at around 2.1~GPa. The higher cell yielding pressure in the cell \#1 is, therefore, due to its stronger inner NiCrAl cylinder in the cell wall. This means that the main source of plastic deformation is at the inner cylinder, which continues to increase with pressure. This was confirmed through material hardness tests, where the inner NiCrAl cylinder used in the cell \#1 had a higher hardness value compared to the cell \#2.

\begin{figure}[t]
\centering
\includegraphics[width=0.75\linewidth]{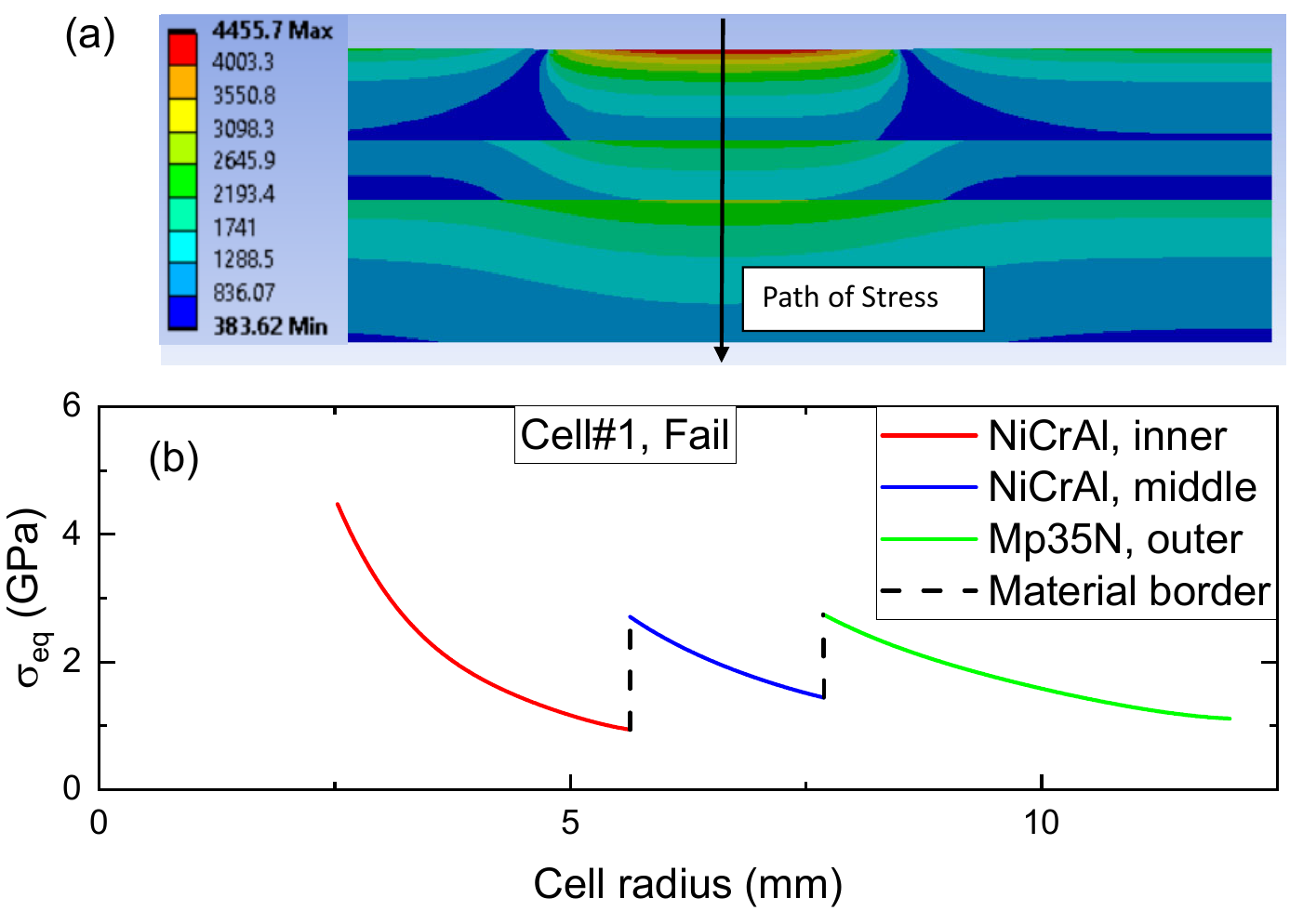}
\caption{(a) Equivalent stress distribution over the cell walls for the pressure cell \#1 at $p=3.79$~GPa. (b) Equivalent stress distribution over of the black pass in the panel (a).}
\label{fig:Cell1_failure}
\end{figure}

Accounting for friction, the cell \#1 was found to fail at an applied load of 74.5~kN, as shown in Fig.~\ref{fig:autofrettage_cell2}~(b) and Table~\ref{Table2}. This equates to $p_{\rm t}=3.79$~GPa, which was applied to the ANSYS FEA model to gain an understanding of the equivalent stress distribution at cell failure [see Fig.~\ref{fig:Cell1_failure}~(a)].
The cell failure source found experimentally was at the inner NiCrAl cylinder. From the FEA results shown in Fig.~\ref{fig:Cell1_failure}~(b), it is clear that the maximum stress at the cell bore is $\simeq 2$~GPa greater than the maximum stress in both the middle and outer NiCrAl and MP35N cylinders. As cell \#1 has not undergone autofrettage, the effective yield strength of the inner cylinders is the same as material yield strength. This means that the middle NiCrAl and outer MP35N cylinders are only around 0.6~GPa and 0.7~GPa over the material yield strength respectively. However, the inner cylinder has a maximum equivalent stress of an almost 4.5~GPa, which is more than double NiCrAl yield strength. The equivalent stress in the inner cylinder propagates above NiCrAl yield strength until approximately 2~mm before the outer cylinder radius. The stress propagation in the inner cylinder will, therefore, allow very little room for the remaining material below yield to contain the high stresses, leading to the failure of the cell.

\section{Scientific example: superconductivity of elemental Indium under pressure}\label{sec:scientific-example}

\subsection{ac susceptibility}

By closing the cell \#2, the pressure inside the cell was determined by measuring the superconducting transition temperature of Indium. Note, that this was the same Indium sample as used for performing the autofrettage cycles (see Sec.~\ref{sec:autofretage_procedure}). The cell was closed at 3.3~GPa by tightening the top bolt and pushing the piston 0.2~mm further down [see Fig.~\ref{fig:autofrettage_cell2}~(a)].

\begin{figure}[t]
\centering
\includegraphics[width=0.55\linewidth]{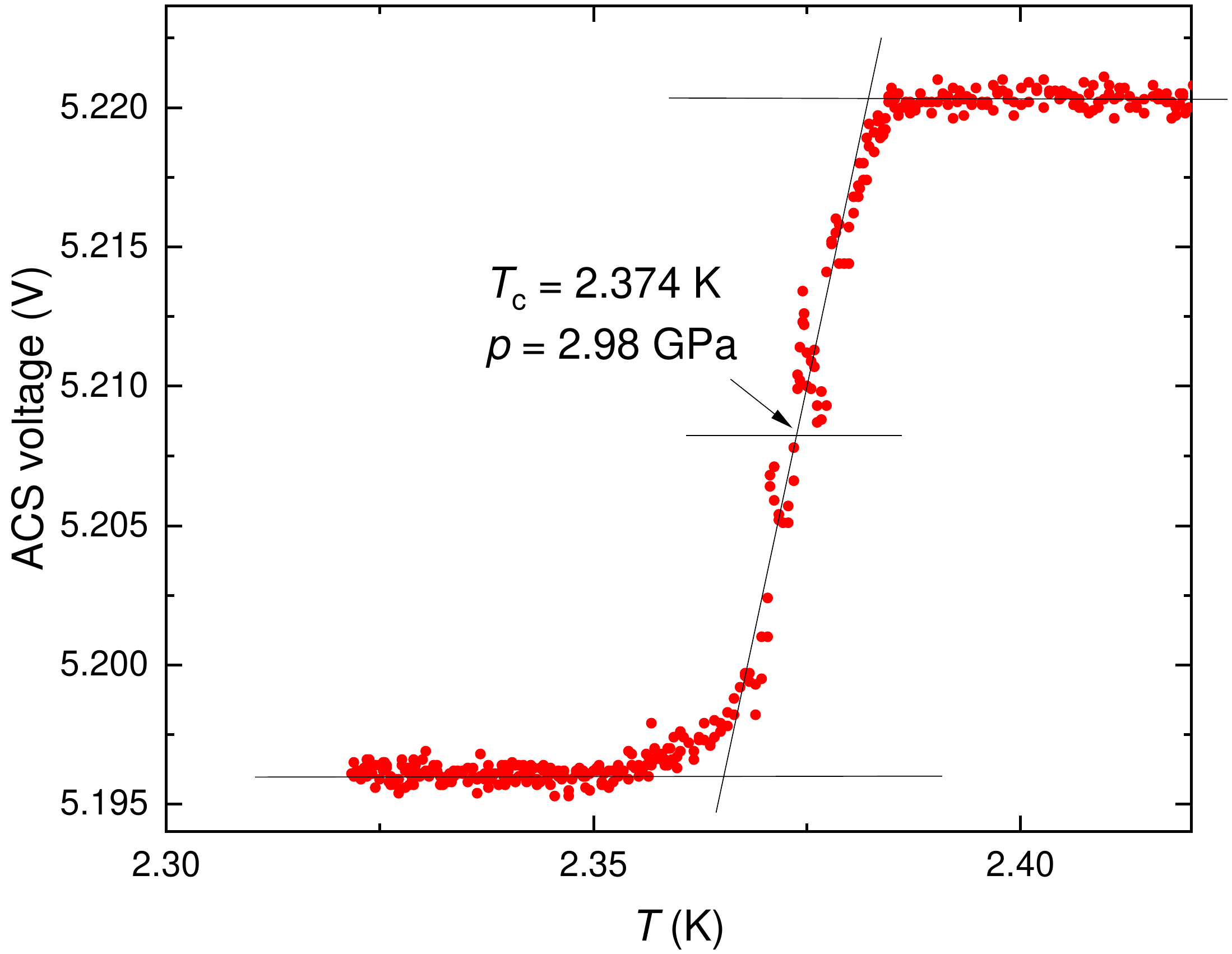}
\caption{The superconducting transition of Indium measured by means of
ac susceptibility. 3.3~GPa pressure is applied at room temperature [see Fig.~\ref{fig:autofrettage_cell2}~(b) and Table~\ref{Table1}]. 2.98~GPa is determined from the pressure induced $T_{\rm c}$ shift of Indium (see Eq.~\ref{eq:Tc_Indium} and Ref.~\cite{Eiling_IJMP_1981}).}
\label{fig:ACS_Indium}
\end{figure}

The superconducting transition of In sample was determined by using the ac susceptibiltiy setup described in Ref.~\cite{Khasanov_HPR_2016}. The pressure inside the cell was then obtained from \cite{Eiling_IJMP_1981}:
\begin{equation}
T_{\rm c}(p) = T_{\rm c}(0)-0.3812\; p + 0.0122\; p^2,
 \label{eq:Tc_Indium}
\end{equation}
where $T_{\rm c}(0)\simeq 3.40$~K is the transition temperature of Indium at $p=0$. The value of $T_{\rm c}({\rm In})=2.374$~K was found, which corresponds to $p\simeq 2.98$~GPa (see Fig.~\ref{fig:ACS_Indium}).

Note that the pressure determined via the transition temperature of In corresponds to the low-temperature value ($T\sim3$~K). The difference between the pressure applied at room temperature (3.3~GPa, see Table~\ref{Table1}) and the low-$T$ pressure is $\simeq 0.3$~GPa, which is typical for piston-cylinder cells and may be caused by different pressure expansion coefficients of the sample, the pressure transmitting media and the material(s) of the pressure cell body  \cite{Khasanov_HPR_2016, Shermadini_HPR_2017, Torikachvili_RSI_2015}.

\subsection{Muon-spin rotation/relaxation experiments}

The muon-spin rotation/relaxation ($\mu$SR) experiments were performed at the $\mu$E1 beamline by using GPD $\mu$SR spectrometer (Paul Scherrer Institute, PSI Villigen, Switzerland) \cite{Khasanov_HPR_2016}. The top-loaded exchange-gas $^4$He cryostat equipped with the $^3$He inset (base temperature $T\simeq 0.25$~K) was used.
The external magnetic field $B_{\rm ex}$ was applied parallel to the direction of the muon momentum and perpendicular to the initial muon-spin direction, which corresponds to the transverse-field (TF) $\mu$SR geometry.

\begin{figure}[t]
\centering
\includegraphics[width=1.0\linewidth]{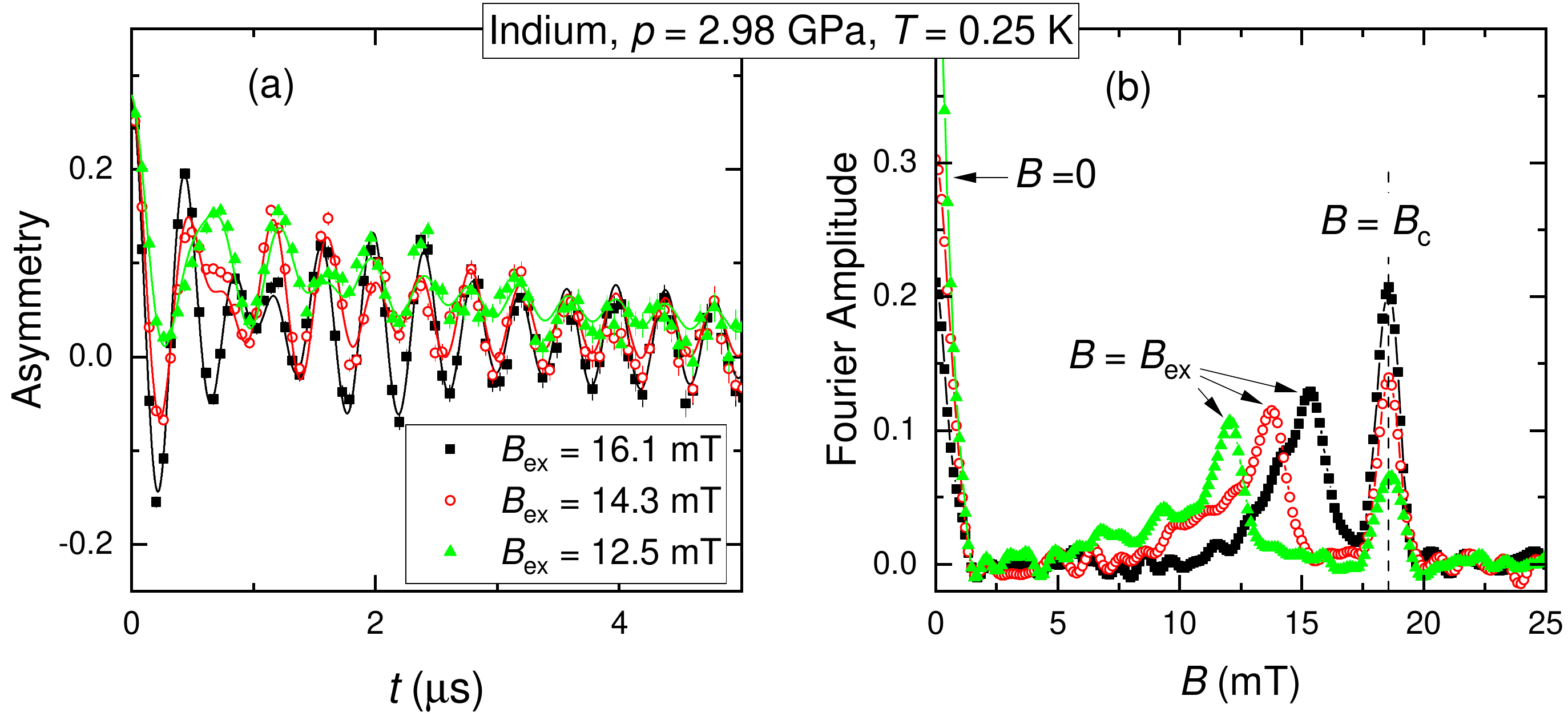}
\caption{ TF-$\mu$SR time spectra collected  on the Indium sample at $T \simeq 0.25$~K, $p=2.98$~GPa, and $B_{\rm ex}=16.1$, 14.3, and 12.5~mT. The solid lines are fits to the experimental data. (b) The Fourier transform of TF-$\mu$SR time spectra presented in the panel (a). Peaks at $B=0$ and $B=B_{\rm c}$ denote the response of the Meissner (superconducting) and the normal state domains, respectively \cite{Khasanov_Bi_PRB_2019, Karl_2019_PRB, Khasanov_Ga_PRB_2020, Khasanov_Pb_PRB_2021}. Peaks at $B=B_{\rm ex}$ represent the background contribution. }
\label{fig:muSR_Indium}
\end{figure}

The raw TF-$\mu$SR data obtained at $T=0.25$~K and $B_{\rm ex}=16.1$, 14.3, and 12.5~mT are shown in Fig.~\ref{fig:muSR_Indium}~(a).  The corresponding Fourier transforms, representing the field distribution inside the In superconductor in the intermediate state, are shown in panel (b).
The experimental data were analyzed by separating the TF-$\mu$SR response of the sample and the background contributions by following the procedure described in Refs.~\cite{Khasanov_Bi_PRB_2019, Karl_2019_PRB, Khasanov_Ga_PRB_2020, Khasanov_Pb_PRB_2021, Khasanov_CrAs_SciRep_2015}.
The magnetic field distribution in a type-I superconductor, as elemental Indium, in the intermediate state, consists of two sharp peaks corresponding to the response of the domains remaining in the Meissner state ($B \equiv 0$) and in the normal state ($B\equiv  B_{\rm c} > B_{\rm ex}$, $B_{\rm c}$ is the thermodynamic critical field), see Fig.~\ref{fig:muSR_Indium}~(b). For the pressurised Indium sample studied in this experiment, the value of $B_{\rm c}(0.25{\rm ~K, 2.98~GPa})=18.55(8)$~mT is obtained, which is almost 10~mT lower compared to the ambient pressure value $B_{\rm c}(0.30{\rm ~K, 0.0~GPa})=28.05$~mT \cite{Finnemore_PRB_1965}.

\section{Summary}\label{sec:conclusions}

To summarize, a three-wall piston–cylinder type hybrid pressure cell was successfully
designed, produced and commissioned. The cell is made out of NiCrAl and
MP35N nonmagnetic alloys with the design and dimensions specifically adapted for
muon-spin rotation/relaxation measurements. For two identical three-wall piston-cylinder high pressure cell designs, the elastic and plastic deformations were modelled using FEA to find the cell expansion during pressurisation and the resulting plastic deformation at the bore. The simulation and experimental results were very similar, where the simulation results were conservative in both cases by only 0.01~mm for cell expansion and between 0.01-0.02~mm for bore diametral plastic deformation. The equivalent stress in each cell at its yield point was also simulated and was found to correlate well with experimental data.

The present design of the three-wall piston–cylinder pressure cell with an outer MP35N sleeve and two inner NiCrAl cylinders allows safely reach pressures  up to $\simeq 3.3$~GPa at room temperature ($\simeq 3.0$~GPa at low temperatures) without irreversible plastic deformation of the pressure cell walls.
Test $\mu$SR experiments were performed on the elemental Indium sample.

\section*{Acknowledgements}
The present work was performed at the Swiss Muon Source (S$\mu$S), Paul Scherrer Institute (PSI, Switzerland). RK thanks to Debarchan Das for taking part in mechanical tests of the pressure cell \#1.

\end{document}